\documentclass[12pt]{article}
\usepackage{graphicx}

\newcommand\pubnumber{SNSN-323-63}
\newcommand\pubdate{\today}

\def\institute{Bergische Universit\"at Wuppertal, 42119 Wuppertal, Germany}
\def\support{\footnote{email: dominic.hirschbuehl@cern.ch}}

\def\Title#1{\begin{center} {\Large #1 } \end{center}}
\def\Author#1{\begin{center}{ \sc #1} \end{center}}
\def\Address#1{\begin{center}{ \it #1} \end{center}}

\newcommand\pubblock{\rightline{\begin{tabular}{l} \pubnumber\\
         \pubdate  \end{tabular}}}
\newenvironment{Abstract}{\begin{quotation}  }{\end{quotation}}
\newenvironment{Presented}{\begin{quotation} \begin{center} 
             PRESENTED AT\end{center}\bigskip 
      \begin{center}\begin{large}}{\end{large}\end{center} \end{quotation}}

\let\bar=\overbar

\def\Dslash{\not{\hbox{\kern-4pt $D$}}}
\def\dslash{\not{\hbox{\kern-2pt $\del$}}}

\def\msb{{\bar{\ssstyle M \kern -1pt S}}}

\begin{document}

\begin{titlepage}
\pubblock

\vfill
\Title{Single-top quark cross-section measurements in ATLAS}
\vfill
\Author{ Dominic Hirschb\"uhl\support \\ on behalf of the ATLAS collaboration}
\Address{\institute}
\vfill
\begin{Abstract}
This article presents measurements of all three single top-quark production
channels. Detailed measurements of $t$-channel single top-quark production using
data collected by the ATLAS experiment in proton--proton collisions at a
centre-of-mass energy of 8 TeV are shown as well es first results using
13 TeV at the LHC.
The associated production of a top quark and a $W$ boson is presented for data
collected at 13 TeV, while the first evidence of single top-quark
production in the $s$-channel is shown for data collected at 8 TeV.
\end{Abstract}
\vfill
\begin{Presented}
$9^{th}$ International Workshop on Top Quark Physics\\
Olomouc, Czech Republic,  September 19--23, 2016
\end{Presented}
\vfill
\end{titlepage}
\def\thefootnote{\fnsymbol{footnote}}
\setcounter{footnote}{0}

\section{Introduction}

At leading order (LO) in perturbation theory, single top-quark production is described by three 
subprocesses that are distinguished by the virtuality of the exchanged $W$ boson. 
The dominant process is the $t$-channel exchange, where a light quark from one
of the colliding protons interacts with a $b$-quark from another proton by exchanging a virtual $W$ boson.
The total inclusive cross-sections of top-quark and top-antiquark production in
the $t$-channel in proton--proton $pp$ collisions at a centre-of-mass
energy $\sqrt{s} = 8$ TeV are predicted to be $\sigma(tq) = 54.9^{+2.3}_{-1.9}$
pb proton--proton $pp$ and $\sigma(\bar{t}q) = 29.7 ^{+1.7}_{-1.5}$ pb for top-antiquark production 
and at $\sqrt{s} = 13$ TeV to be $\sigma(tq)= 136.0^{+5.4}_{-4.6}\;$pb  and
$\sigma(\bar{t}q)=81.0^{+4.1}_{-3.6}\;$pb at next-to-leading order (NLO)
precision in QCD~\cite{Campbell:2009ss,Kant:2014oha}.
The second highest production cross section is predicted for the associated
production of a $W$ boson and a top quark~($Wt$).
The cross-section of the $Wt$ channel at NLO with next-to-next-to-leading logarithmic 
soft-gluon corrections is calculated as
$\sigma(Wt)= 71.7 \pm 3.8\;$pb~\cite{Kidonakis:2015nna} for $\sqrt{s} = 13$ TeV.
The $s$-channel production of a top quark and a $b$-quark ($t\bar{b}$) for
$\sqrt{s} = 8$ yields the smallest cross section
of $\sigma(s)= 5.61 \pm 0.22$ pb in NLO QCD~\cite{Stelzer:1998ni}.\\
In the following, measurements of all three production channels
using data collected by the ATLAS experiment~\cite{Aad:2008zzm} in
$pp$ collisions either at $\sqrt{s} = 8$ TeV
corresponding to an integrated luminosity of 20.2 fb$^{-1}$
or $\sqrt{s} = 13$ TeV at the LHC corresponding to an integrated luminosity of
3.2 fb$^{-1}$ are presented.

\section{$s$-channel single top-quark production}
\begin{figure}[htb]
\centering
\includegraphics[width=0.42\textwidth]{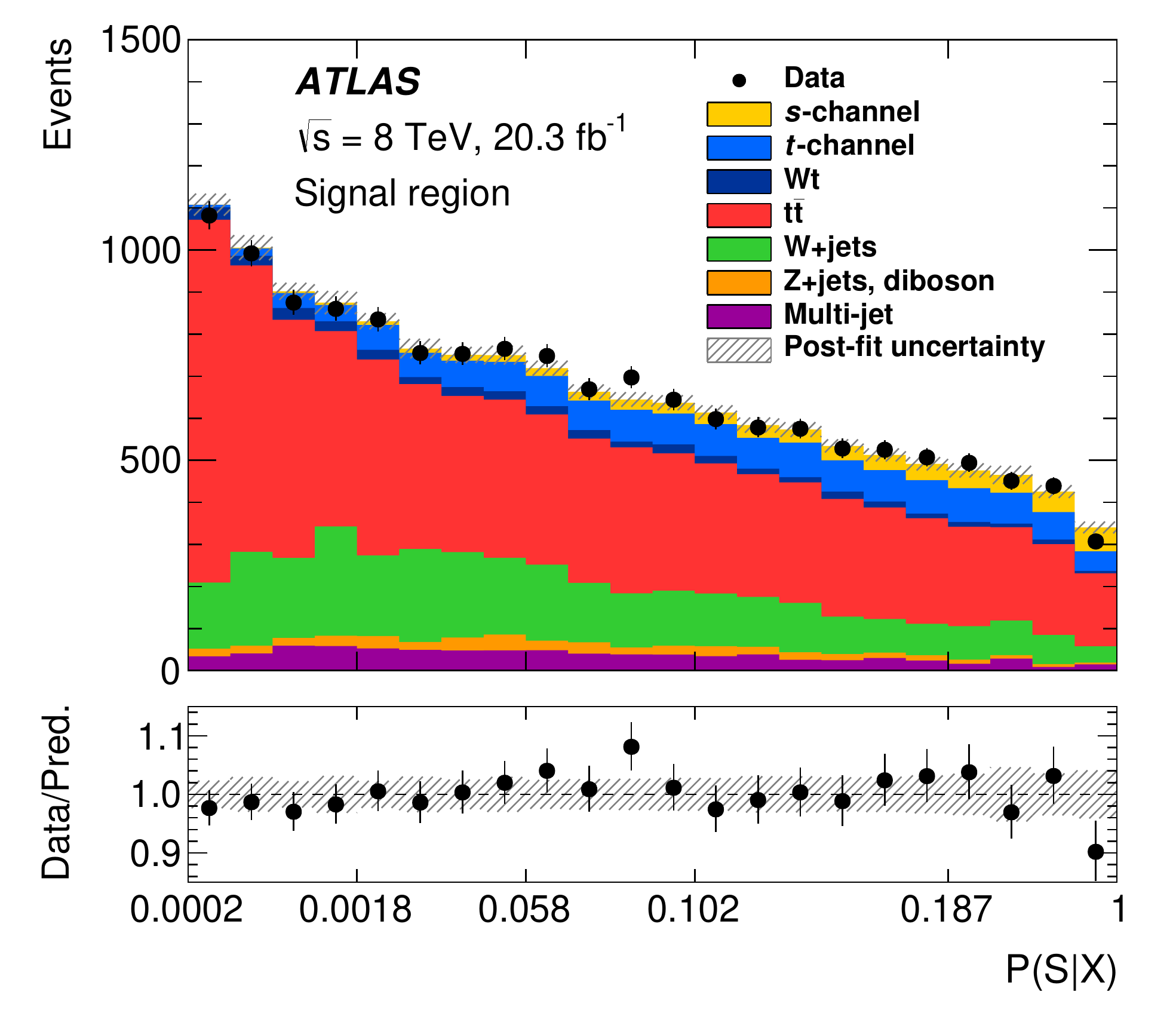}
\includegraphics[width=0.54\textwidth]{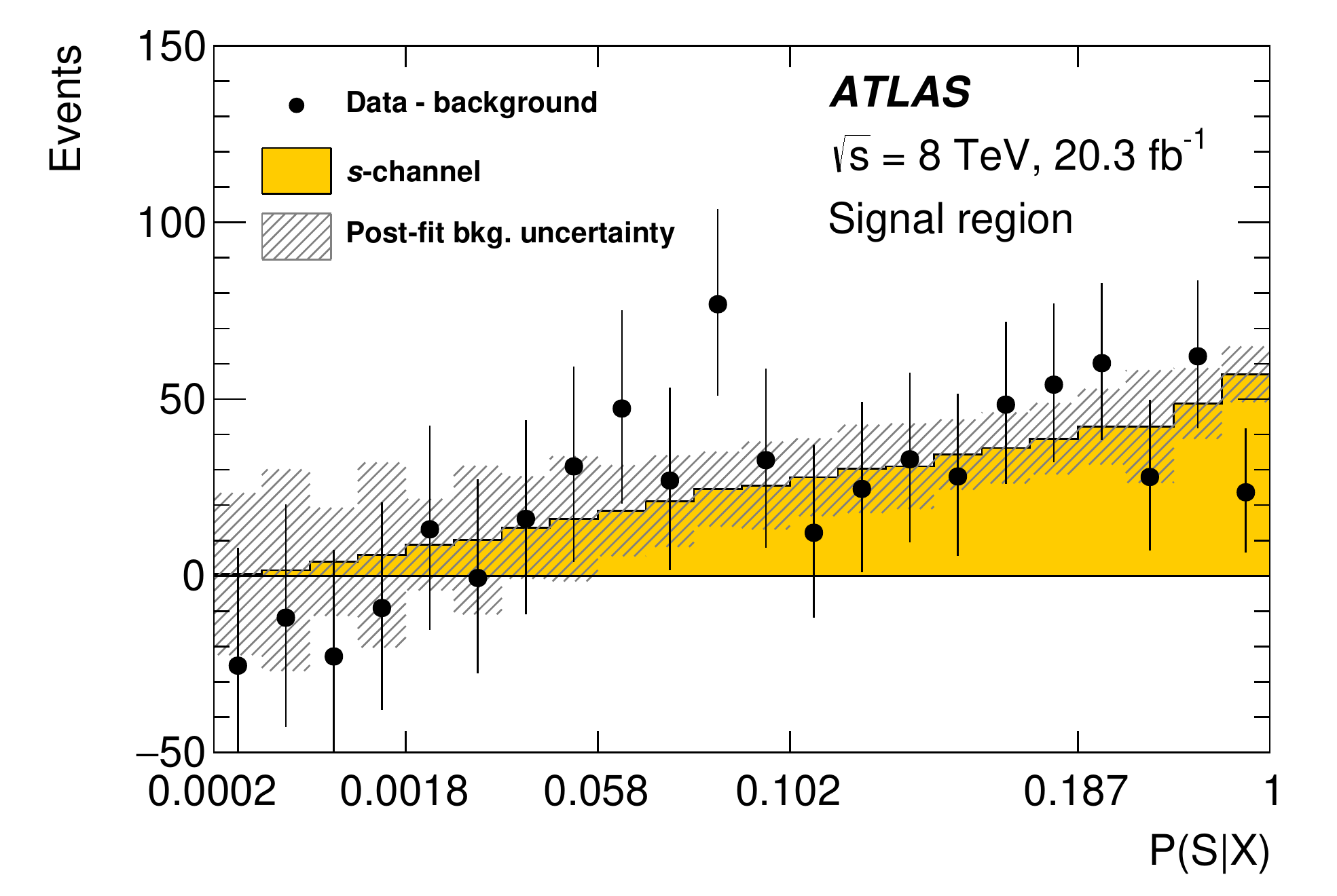}
\caption{Post-fit distribution of the ME discriminant in the signal region
         (left). The hatched bands indicate the total uncertainty of the post-fit result including all correlations. 
         Distribution of the ME discriminant in data in the signal region after
         the subtraction of all backgrounds (right), showing the signal
         contribution. The error bars indicate the uncertainty of the
         measurement in each bin~\cite{Aad:2015upn}.}
         
\label{fig:mm}
\end{figure}

The analysis for $s$-channel single top-quark production 
is performed using collision data at $\sqrt{s} = 8$ TeV.
Events are selected if they have either of an electron or muon, two jets,
where both have to be identified as a jet containing $b$-hadrons ($b$-tagged
jet) and large missing transverse momentum $E_{\mathrm{T}}^{\mathrm{miss}}$. 
After the preselection, the main background are top-quark pair-production
($t\bar{t}$) and $W$+jets production. In order to
separate the signal from the large background contributions, a matrix element method discriminant is
used, see Fig.~\ref{fig:mm}. 
The signal is extracted from the data utilising a
profile likelihood fit, which leads to a measured cross-section of
$\sigma(s) = 4.8 \pm 0.8 \mathrm{(stat)}^{+1.6}_{-1.3} \mathrm{(syst)}$
pb~\cite{Aad:2015upn}.
Dominating uncertainties are Monte Carlos (MC) statistics, jet energy
resolution, and the modelling of the $t$-channel single top-quark process.
The result, which is in agreement with the SM prediction, corresponds to an observed significance of 3.2
standard deviations.

\section{Associated Wt production}
\begin{figure}[htb]
\centering
\includegraphics[width=0.75\textwidth]{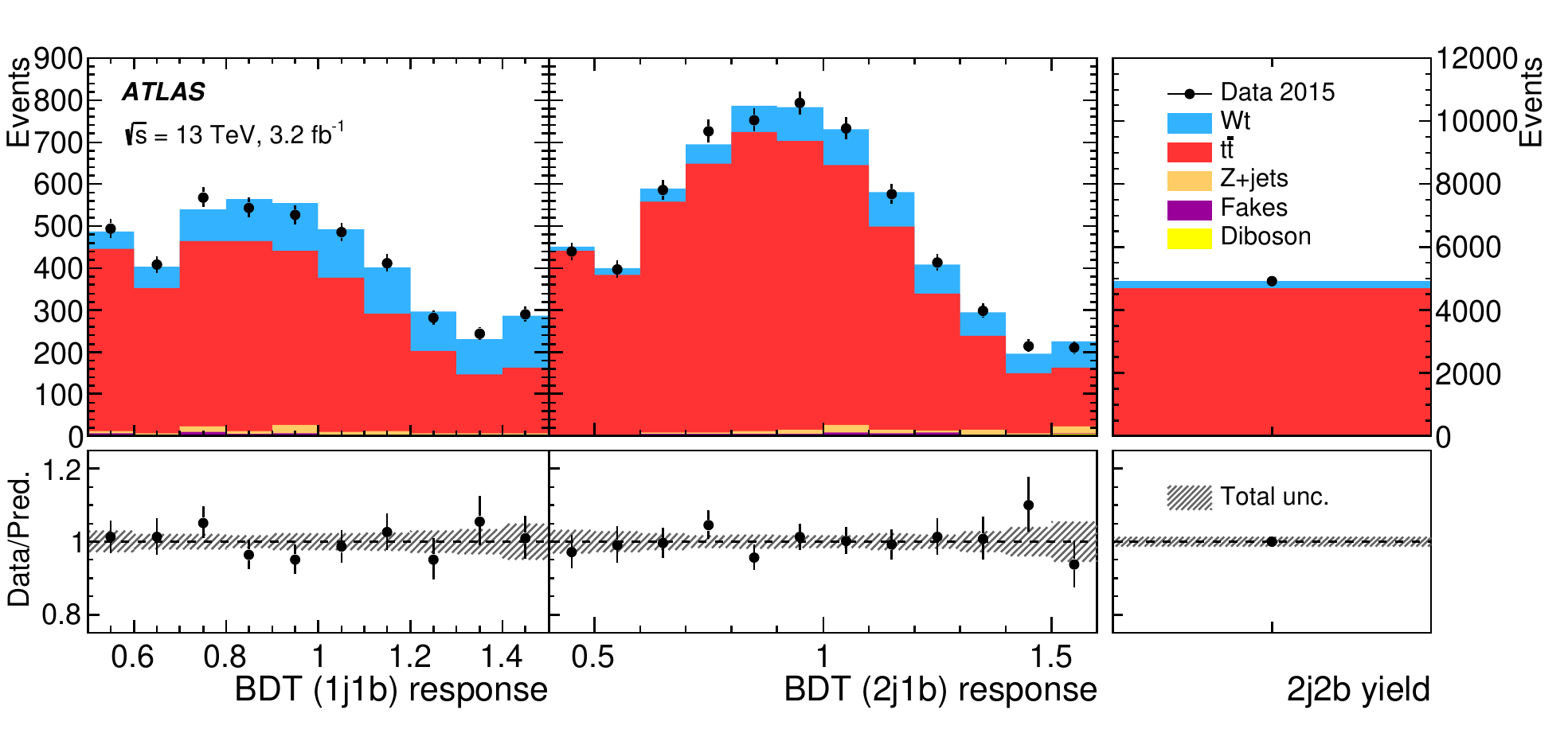}

\caption{Post-fit distributions of the signal and control regions.
         The error bands represent the total uncertainties on the
         fitted results. The upper panels give the yields in number of events
         per bin, while the lower panels give the ratios of the numbers of
        observed events to the total prediction in each
        bin~\cite{Aaboud:2016lpj}. }
\label{fig:Wt}
\end{figure}

The inclusive cross-section for the associated production of a $W$ boson and top quark is measured 
using dilepton events with at least one $b$-tagged jet.
Events are separated into signal and control regions based on the number of
selected jets and $b$-tagged jets, and the $Wt$ signal is separated from the
$t\bar{t}$ background using a boosted decision tree (BDT) discriminant,
shown in Fig.~\ref{fig:Wt} The cross-section is extracted by fitting templates
to the BDT output distribution, and is measured to be $\sigma(Wt) = 94 \pm 10 \mathrm{(stat)} ^{+28}_{-23}
\mathrm{(syst)}$ pb~\cite{Aaboud:2016lpj}. 
Main uncertainties are coming from the jet energy scale (JES) and the modelling
of the top-quark processes.

\section{$t$-channel single top-quark production}
\begin{figure}[htb]
\centering
\includegraphics[width=0.455\textwidth]{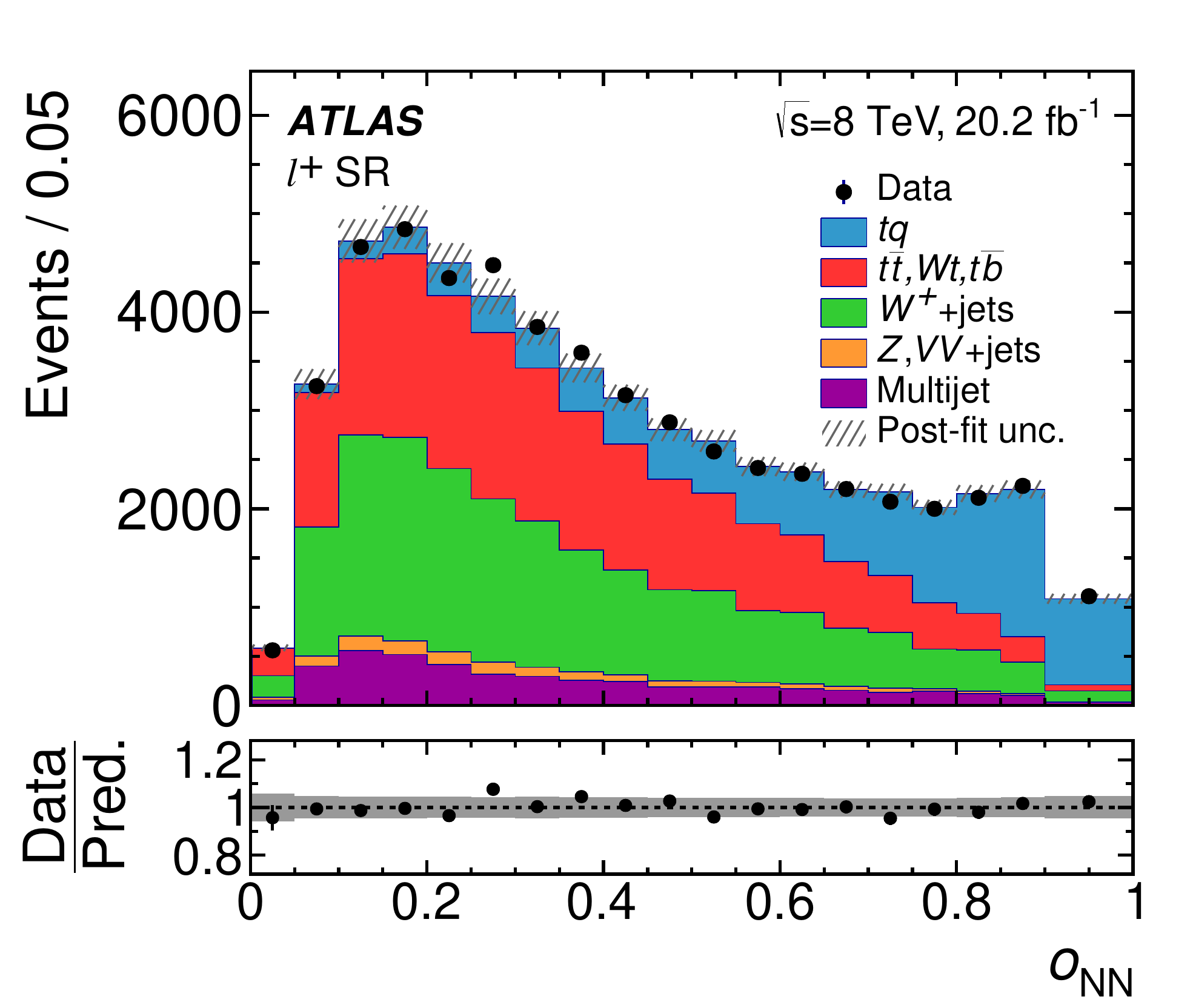}
\includegraphics[width=0.455\textwidth]{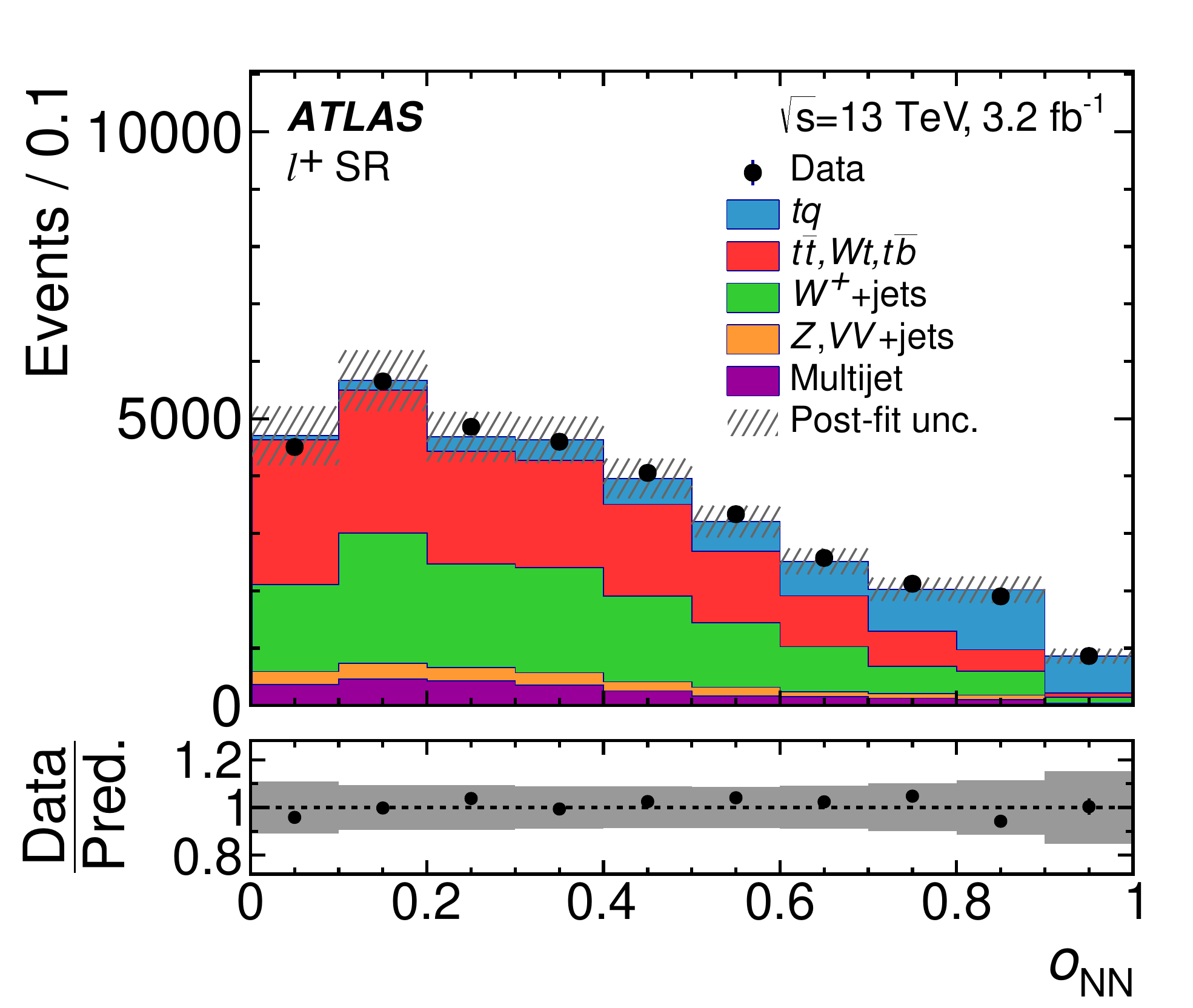}
\caption{NN discriminant distribution for positively charged leptons in
         the SR (left) for 8 TeV~\cite{Aaboud:2017pdi} and (right) for 13
         TeV~\cite{Aaboud:2016ymp}.
  The signal and backgrounds are normalised to the fit result and 
  the hatched and grey error bands represent the post-fit uncertainty.
  The ratio of observed to predicted number of events in each bin is shown in the lower histogram.}
\label{fig:nn}
\end{figure}

\begin{figure}[htb]
\centering
\includegraphics[width=0.455\textwidth]{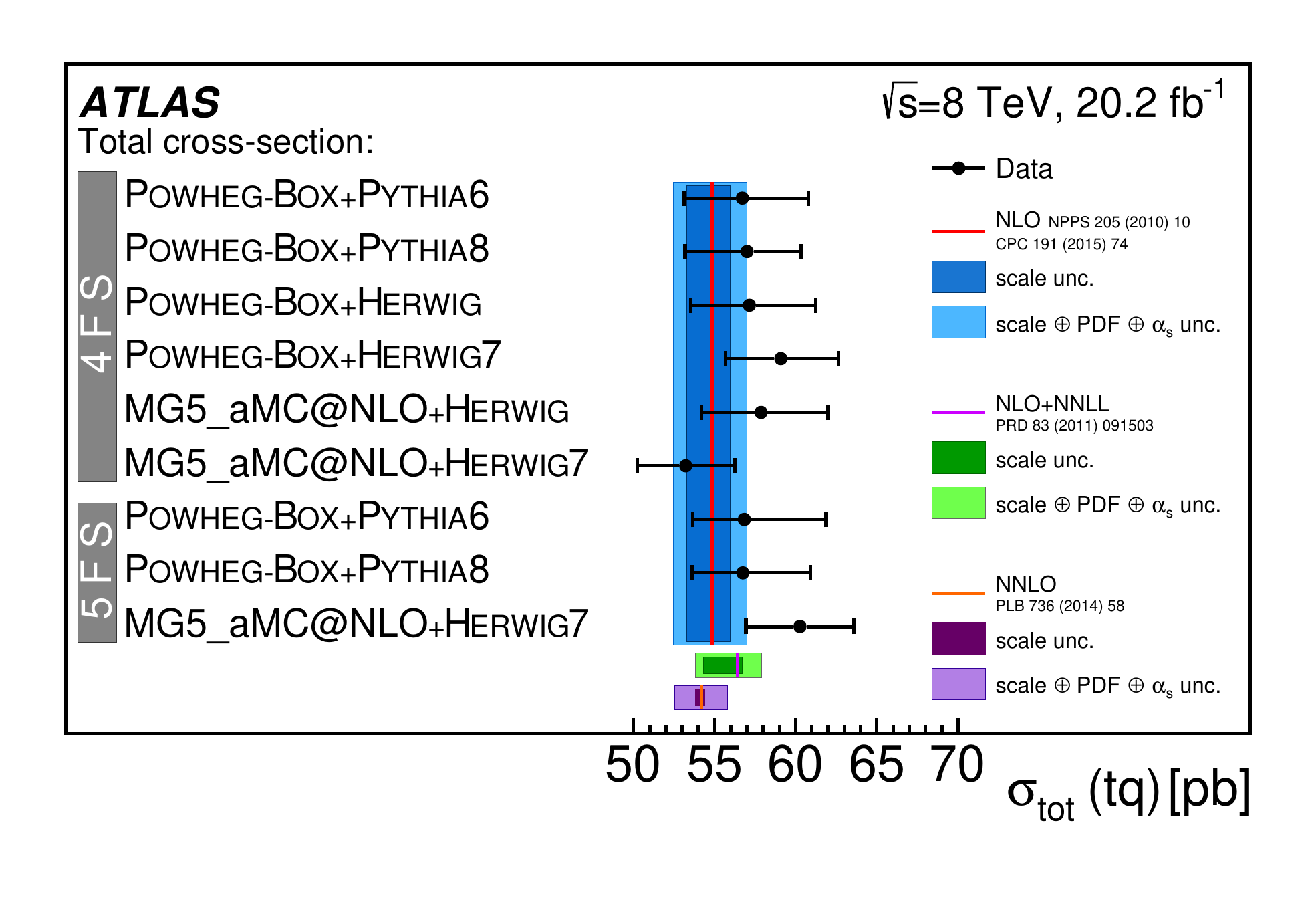}
\includegraphics[width=0.455\textwidth]{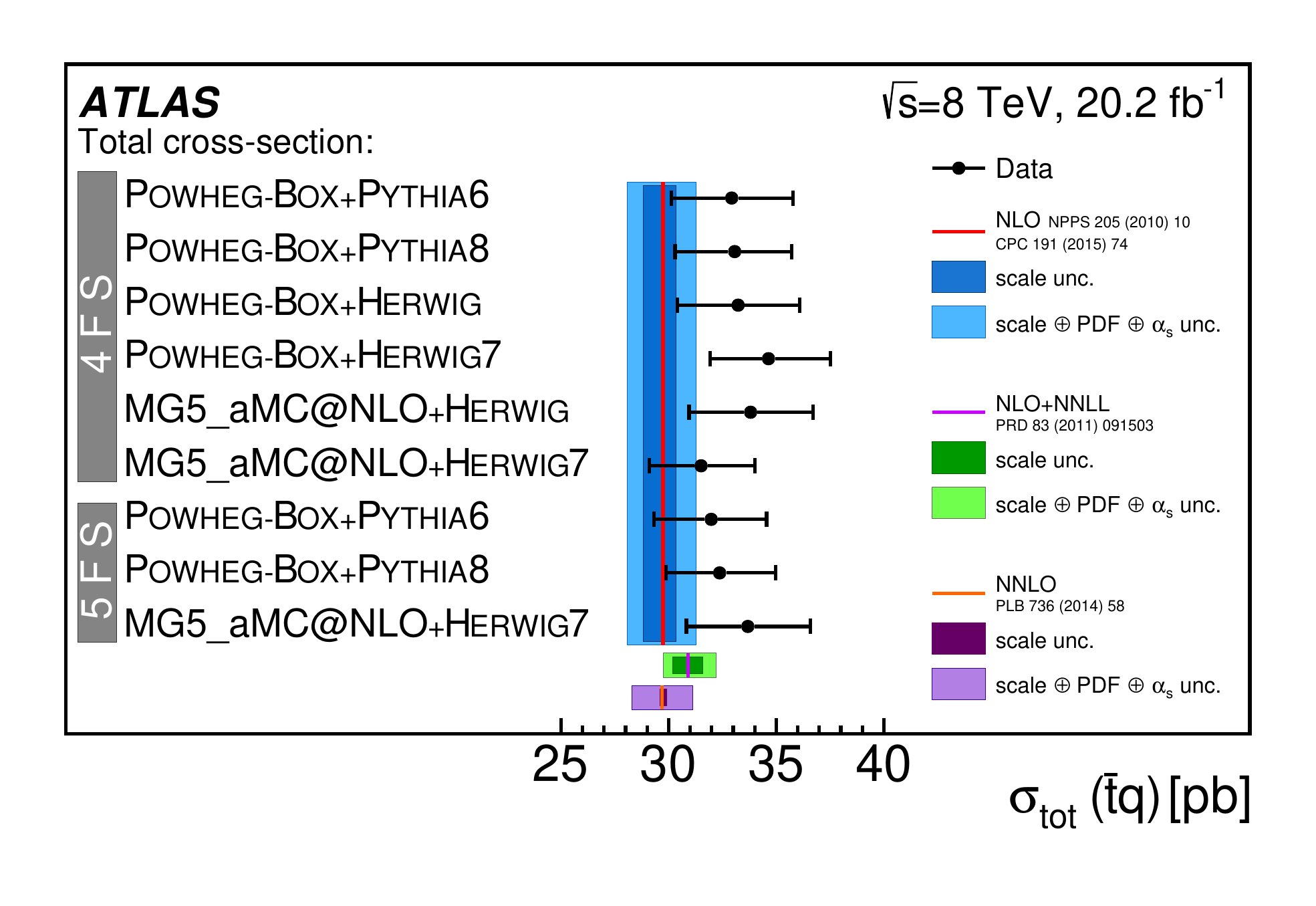}
\caption{Measured $t$-channel single-top-quark and
single-top-antiquark fiducial cross-sections compared to predictions by the NLO
MC generators \textsc{POWHEGBOX} and \textsc{MG5\_aMC@NLO} in the four-flavour
scheme (4FS) and five-flavour scheme (5FS) combined with different parton-shower models
(left) for 8 TeV~\cite{Aaboud:2017pdi} and (right) for 13
TeV~\cite{Aaboud:2016ymp}.
The uncertainties in the predictions include the uncertainty due to the scale
choice and the intra-PDF uncertainty.}
\label{fig:fiducial}
\end{figure}

\begin{figure}[htb]
\centering
\includegraphics[width=0.455\textwidth]{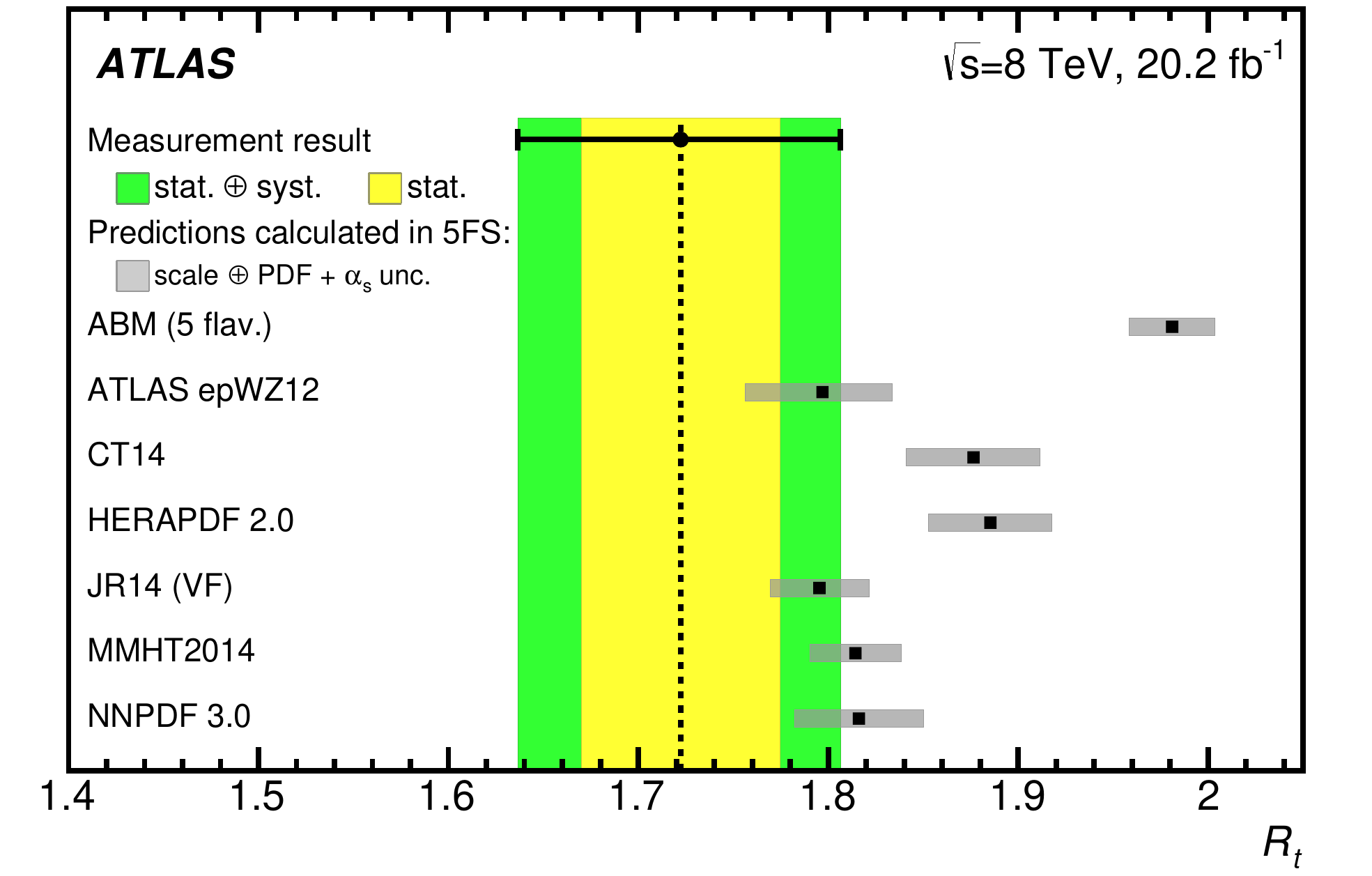}
\includegraphics[width=0.455\textwidth]{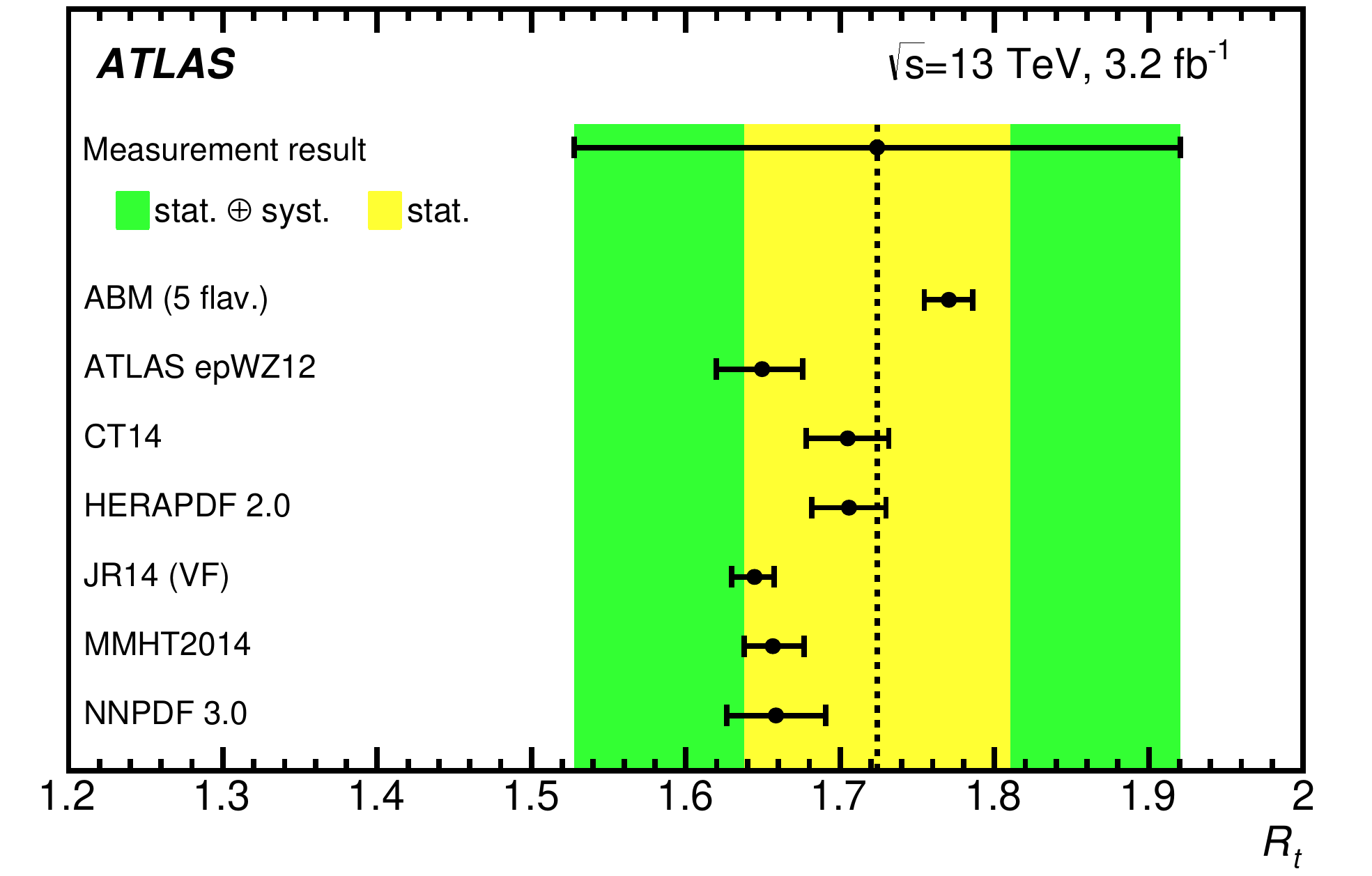}
\caption{Comparison between observed and predicted values of
$R_t=\frac{\sigma_{t}}{\sigma_{\bar{t}}}$ (left) for 8 TeV~\cite{Aaboud:2017pdi}
and (right) for 13 TeV~\cite{Aaboud:2016ymp}. Predictions are calculated at NLO
 precision~\cite{Campbell:2009ss,Kant:2014oha} in the five-flavour scheme and given for different NLO PDF sets. The uncertainty includes the uncertainty in the renormalisation and factorisation scales, as well as the combined internal PDF and $\alpha_{\rm s}$ uncertainty.
  The dotted black line indicates the measured value. 
  The combined statistical and systematic uncertainty of the measurement is shown in green, 
  while the statistical uncertainty is represented by the yellow error band.}
\label{fig:rt}
\end{figure}

The experimental signature of $t$-channel single top-quark candidate events is
given by one charged lepton (electron or muon), large
$E_{\mathrm{T}}^{\mathrm{miss}}$ and two jets.
Exactly one of the two jets is required to be $b$-tagged.
The main background contributions are the $t\bar{t}$ and $W$+jets processes.
In order to separate signal from background an artificial neural network is
used, shown in Fig.~\ref{fig:nn}\\
For the $\sqrt{s} = 8$ TeV data; the total and fiducial cross-sections are
measured for both top quark and top antiquark production.
The fiducial cross-section is measured with a precision of 5.8\%
(top quark) and 7.8\% (top antiquark), respectively~\cite{Aaboud:2017pdi}.
A comparison with different MC generator setups is shown in
Fig.~\ref{fig:fiducial} for the extrapolated total cross section. 
In addition, the cross-section ratio of top-quark to top-antiquark production
is measured, resulting in a precise value to compare with predictions,
$R_t =\frac{\sigma_{t}}{\sigma_{\bar{t}}} = 1.72 \pm 0.09$ and presented in
Fig.~\ref{fig:rt}(left).
Dominant uncertainties for these measurements are the JES and modelling of the
top-quark processes.
The total cross-section is used to extract
the $Wtb$ coupling: $f_{\mathrm{LV}} \cdot |V_{tb}| = 1.029\pm 0.048$,
which corresponds to $|V_{tb}|>0.92$ at the 95\% confidence level,
when assuming $f_{\mathrm{LV}} = 1$ and restricting the range of $|V_{tb}|$ to
the interval $[0, 1]$.\\
Requiring a high value of the neural-network discriminant leads to relatively
pure $t$-channel samples, which are used to measure differential cross-sections.
Differential cross-sections as a function of the transverse momentum and
absolute value of the rapidity of the top quark, the top antiquark,
as well as the accompanying jet from the $t$-channel scattering are
measured at particle level and parton level.
All measurements are compared to different Monte Carlo predictions as well as
to fixed-order QCD calculations where these are available.
The SM predictions provide good descriptions of the data.\\
For the $\sqrt{s} = 13$ TeV data; the the total 
cross-sections for both top quark and top antiquark production are measured to
be $\sigma(tq) = 156 \pm 28 ~\mathrm{pb}$ and $\sigma(\bar{t}q) = 91 \pm 19
~\mathrm{pb}$, respectively~\cite{Aaboud:2016ymp}.\\
The cross-section ratio is found to be $R_t = 1.72 \pm 0.20$ and compared with
predictions from different PDF groups in Fig.~\ref{fig:rt} (right).
The coupling at the $Wtb$ vertex is determined to be $\mathrm{LV} \cdot |V_{tb}|=1.07 \pm 0.09$ and
a lower limit on the CKM matrix element is set, giving $|V_{tb}| > 0.84$ at the
95\% CL.
These measurements are dominated by systematic uncertainties,
from which the uncertainties connected with MC generators are the biggest ones.

\end{document}